\documentclass{article}
\usepackage{amsfonts}
\usepackage{natbib}
\usepackage{colonequals} 
\usepackage{amsmath}
\usepackage{amssymb,epsfig,color,multirow}
\usepackage{subfigure}

\AtEndOfClass{\RequirePackage{times}}
\newcommand{\sampcov}{\mathbf{S}}
\newcommand{\ident}{\mathbf{I}}

\newcommand{\vecx}{\mathbf{x}}
\newcommand{\vecv}{\mathbf{v}}

\newcommand{\vecz}{\mathbf{z}}

\newcommand{\varthet}{\mbox{\boldmath$\vartheta$}}
\newcommand{\vecmu}{\mbox{\boldmath$\mu$}}

\newcommand{\matd}{\mathbf{D}}
\newcommand{\mata}{\mathbf{A}}
\newcommand{\matb}{\mathbf{B}}
\newcommand{\matp}{\mathbf{P}}
\newcommand{\mats}{\mathbf{S}}

\newcommand{\matq}{\mathbf{Q}}

\newcommand{\matsig}{\mathbf\Sigma}

\begin{document}

\title{Constrained Optimization for a Subset of the \\ Gaussian Parsimonious Clustering Models}
\author{Ryan P.\ Browne\thanks{Department of Mathematics \& Statistics, University of Guelph, Guelph, Ontario, N1G 2W1, Canada. E-mail: {\tt rbrowne@uoguelph.ca}}, \ Sanjeena Subedi and Paul D. McNicholas}
\date{Department of Mathematics and Statistics, University of Guelph, Ontario, Canada.}
\maketitle

\begin{abstract}
The expectation-maximization (EM) algorithm is an iterative method for finding maximum likelihood estimates when data are incomplete or are treated as being incomplete. The EM algorithm and its variants are commonly used for parameter estimation in applications of mixture models for clustering and classification. This despite the fact that even the Gaussian mixture model likelihood surface contains many local maxima and is singularity riddled. Previous work has focused on circumventing this problem by constraining the smallest eigenvalue of the component covariance matrices. In this paper, we consider constraining the smallest eigenvalue, the largest eigenvalue, and both the smallest and largest within the family setting. Specifically, a subset of the GPCM family is considered for model-based clustering, where we use a re-parameterized version of the famous eigenvalue decomposition of the component covariance matrices. Our approach is illustrated using various experiments with simulated and real data.
\end{abstract}

\section{Introduction}
The expectation-maximization (EM) algorithm \citep{dempster77} is an iterative procedure for finding maximum likelihood estimates when data are incomplete or treated as such. Although the EM algorithm is commonly attributed to \cite{dempster77}, \citet[][Section~4.3.2]{titterington85} point out that similar treatments had previously been employed by \cite{baum70}, \cite{orchard72}, and \cite{sundberg74}. The EM algorithm involves the iteration of two steps until convergence is attained. In the expectation step (E-step) the expected value of the complete-data log-likelihood is computed and then, in the maximization step (M-step), this expected value is maximized with respect to the model parameters. Here, `complete-data' refers to the missing plus observed data.

In this paper, we are concerned with the application of the EM algorithm to Gaussian mixture models with applications in clustering and classification. The likelihood surface for Gaussian mixture models is known to be unbounded and the presence of local maxima is extremely common; one may go so far as to argue that the surface is singularity riddled \citep{titterington85}.
When using the EM algorithm to fit Gaussian mixture models, problems like convergence to a spurious local maxima tend to arise when one fitted component has much smaller variance than the others \citep[cf.][]{biernacki04}; an illustrative example of this phenomenon is given by \cite{ingrassia2007}, who we will follow by referring to such fitted components as `degenerate'. The behaviour of the EM algorithm near a degenerate solution has been studied by \cite{biernacki03} and \cite{ingrassia2007}, who tackle the problem by constraining the value of the smallest eigenvalue of the component covariance matrices. In this paper, we consider constraining the smallest eigenvalue, the largest eigenvalue, and both the smallest and largest eigenvalues. While this approach is applicable to Gaussian mixture models in general, we impose these constraints  while maintaining a parsimonious covariance structure. We focus on a 
subset of the famous parsimonious Gaussian clustering models (GPCM) family of mixture models of \cite{celeux95}, cf.\ Section~\ref{sec:back}.


This paper illustrates the benefit of constraining the range (minimum and maximum) of eigenvalues in two applications. In what we call `dynamic initialization', we begin with at a random starting value and impose stringent constraints on the eigenvalues and these constraints are slowly lifted during the first $k$ iterations of the EM algorithm. Dynamic initialization maintains the monotonicity property of the EM algorithm while reducing the risk of converging  to a degeneracy. We also show that in most cases, dynamic initialization increases the changes of converging to a solution with higher log-likelihood than when compared to using the `standard' (with no dynamic initialization) EM algorithm with the same starting values. 
The other application of constraining the range of eigenvalues is a more direct one. We use a constraint on the range of eigenvalues to fit the relevant GPCM models to two well known data sets within the model-based clustering literature. For each data set, we find solutions that are an improvement over the GPCM models; however, we must choosing constraints \emph{a priori} (cf.\ Section~\ref{sec:conc}).

The remainder of the paper is laid out as follows. In Section~\ref{sec:back}, the GPCM family of models is introduced. Then parameter estimation while constraining the largest and/or smallest eigenvalue is discussed and our methodology for constraining both is described (Section~\ref{sec:method}). Our approach is illustrated via several experiments using real and simulated data (Section~\ref{sec:anal}) and the paper concludes with discussion and suggestions for future work (Section~\ref{sec:conc}). 


\section{Background}
\subsection{The GPCM and MCLUST families}\label{sec:back}
Suppose that we observe $n$ $p$-dimensional data vectors $\vecx_1,\ldots,\vecx_n$ and that each must be assigned to one of $G$ clusters. The Gaussian mixture model-based approach has become popular for such problems. When mixture models are used for clustering in this way, the locution `model-based clustering' is used. The Gaussian model-based clustering likelihood is 
$$\mathcal{L}(\varthet\mid\vecx_1,\ldots,\vecx_n)=\prod_{i=1}^n\sum_{g=1}^G\pi_g\phi(\vecx_i\mid\vecmu_g,\matsig_g),$$
where  $\pi_g >0$, such that $\sum_{g=1}^G \pi_g = 1$, are mixing proportions and $\phi(\vecx_i\mid\vecmu_g,\matsig_g)$ is the density of a multivariate Gaussian random variable with mean~$\vecmu_g$ and covariance matrix $\matsig_g$. 
Extensive details on finite mixture models and their applications are given by \cite{titterington85}, \cite{mclachlan88}, \cite{mclachlan00b}, and \cite{fruhwirth06}.

Unless $p$ is small relative to $n$, model fitting issues arise with these models because of the large number of covariance parameters; there are $p(p+1)/2$ parameters for each component covariance matrix $\matsig_g$. \cite{celeux95} introduce parsimony into these Gaussian mixture models by proposing and giving estimation algorithms for fourteen different eigen-decompositions of the covariance matrix (Table~\ref{tab:mclust2}). These decompositions have the form $\matsig_g=\lambda_g\matd_g\mata_g\matd_g'$, where $\matd_g$ is the matrix of eigenvectors, $\mata_g$ is a diagonal matrix with entries proportional to the eigenvalues, and $\lambda_g$ is the associated constant of proportionality. The resulting 14 models are called Gaussian parsimonious clustering models (GPCMs). \cite{fraley98} implemented ten of these fourteen models, based on the algorithms given in \cite{celeux95}, as the popular {\tt mclust} package for the {\sf R} software \citep{R13}. Such has been the popularity of the MCLUST package \citep{fraley12} that only ten of the fourteen GPCMs are routinely employed. \cite{browne13} implemented all fourteen models in the {\tt mixture} package for the {\sf R} software, using the algorithms given in \cite{celeux95} and \cite{browne13b}. 
\begin{table}[h]
\caption{\label{tab:mclust2}Nomenclature, covariance structure, and number of free covariance parameters for each member of the GPCM family; all models are available within {\tt mixture} whereas the last four are not included within {\tt mclust}.} 
{\scriptsize\centering
\begin{tabular*}{1.0\textwidth}{@{\extracolsep{\fill}}lllllr}
\hline
Mod. &  Volume & Shape & Orient. & $\matsig_g$ & Free covariance parameters\\
\hline
EII & Equal & Spherical & --& $\lambda\ident$ & 1\\
VII & Variable & Spherical & -- & $\lambda_g\ident$ & $G$\\
EEI & Equal & Equal & Axis-Aligned & $\lambda\mata$ & $p$\\
VEI & Variable & Equal & Axis-Aligned & $\lambda_g\mata$ & $p+G-1$\\
EVI & Equal & Variable & Axis-Aligned & $\lambda\mata_g$ & $pG-G+1$\\
VVI & Variable & Variable & Axis-Aligned & $\lambda_g\mata_g$ & $pG$\\
EEE & Equal & Equal & Equal & $\lambda\matd\mata\matd'$ & $p(p+1)/2$\\
EEV & Equal & Equal & Variable & $\lambda\matd_g\mata\matd_g'$ & $Gp(p+1)/2 - (G-1)p$\\
VEV & Variable & Equal & Variable & $\lambda_g\matd_g\mata\matd_g'$ & $Gp(p+1)/2 - (G-1)(p-1)$\\
VVV & Variable & Variable & Variable & $\lambda_g\matd_g\mata_g\matd_g'$ & $Gp(p+1)/2$\\
EVE & Equal & Variable & Equal & $\lambda\matd\mata_g\matd'$ & $p(p+1)/2 +(G-1)(p-1)$\\
VVE & Variable & Variable & Equal & $\lambda_g\matd\mata_g\matd'$ & $p(p+1)/2 + (G-1)p$\\
VEE & Variable & Equal & Equal & $\lambda_g\matd\mata\matd'$ & $p(p+1)/2 + (G-1)$\\
EVV & Equal & Variable & Variable & $\lambda\matd_g\mata_g\matd_g'$ & $Gp(p+1)/2-(G-1)$\\
\hline
\end{tabular*}}
\end{table}

In this paper, we consider a reduced set of GPCMs, that we refer to as the rGPCM family 
%
(cf.\ Table~\ref{tab:mclust3}). We are, in effect, reparameterizing the GPCM covariance structure to two parameters by writing $\matb_g = \lambda_g \mata_g$ and $\matb = \lambda\mata$, where $\matb_g$ and $\matb$ are unconstrained matrices of eigenvalues. Not including $\lambda_g$ as a parameter ties together component volume and shape in terms of whether they are constrained, i.e., $\matb_g = \lambda_g \mata_g$ and $\matb = \lambda\mata$, while allowing component orientation $\matd_g$ to vary separately. One may argue that the paramterization $\matsig_g=\matd_g\matb\matd_g'$ is attractive because it has a `natural' eigenvalue interpretation; however, it inherently has less flexibility than the GPCM models (i.e., $\matsig_g=\lambda\matd_g\mata\matd_g'$) and this loss of flexibility needs to be considered in context with the benefits of our constrained eigenvalue decomposition.
\begin{table}[h]
\caption{\label{tab:mclust3}Nomenclature, covariance structure, and number of free covariance parameters for each member of the rGPCM family; equivalent models are available within {\tt mixture} but no equivalent for the last model is available within {\tt mclust}.} 
{\scriptsize\centering\begin{tabular*}{1.0\textwidth}{@{\extracolsep{\fill}}llllr}
\hline
Model &  Volume/Shape & Orientation & $\matsig_g$ & Free covariance parameters\\
\hline
1I & Equal & - & $\lambda\ident$ & $1$\\
GI & Variable & - & $\lambda_g\ident$ & $G$ \\
EI & Equal & Axis-Aligned & $\lambda\mata$ & $p$\\
VI & Variable & Axis-Aligned & $\lambda_g\mata_g$ & $pG$\\
EE & Equal & Equal & $\lambda\matd\mata\matd'$ & $p(p+1)/2$\\
EV & Equal & Variable & $\lambda\matd_g\mata\matd_g'$ & $Gp(p+1)/2 - (G-1)p$\\
VV & Variable & Variable & $\lambda_g\matd_g\mata_g\matd_g'$ & $Gp(p+1)/2$\\
VE & Variable & Equal & $\lambda_g\matd\mata_g\matd'$ & $p(p+1)/2 + (G-1)p$\\
\hline
\end{tabular*}}
\end{table}

We develop an EM algorithm to fit the rGPCM models while imposing constraints on the eigenvalues. In addition, we investigate algorithms that slowly lift the constraints on the eigenvalues either from below, above, or both. \cite{ingrassia2004} shows that having lower and upper bounds on the eigenvalues does not destroy the monotonicity property of the EM algorithm. 
Furthermore, keeping the eigenvalues from going below a threshold prevents degeneracy of the log-likelihood \citep{ingrassia2011}. However, algorithms that only have dynamic constraints from below prolong degeneracy.  We show that having dynamic constraints from above and below reduces the risk of degeneracy and yields higher log-likelihood values at convergence. 

\subsection{Parameter Estimation and Model Selection}\label{sec:para}
%

Parameter estimation for each member of the GPCM family is carried out using an EM algorithm. The EM algorithm is used to obtain maximum likelihood estimates when data are incomplete or are taken to be incomplete. In mixture model-based clustering applications, the missing data are the component membership labels, which we denote by $\vecz_1,\ldots,\vecz_n$, where $z_{ig}=1$ if observation~$i$ is in component~$g$ and $z_{ig}=0$ otherwise. These missing data together with the observed data $\vecx_1,\ldots,\vecx_n$ are known as the complete-data, and the E-step of the EM algorithm involves computation of the expected value of the complete-data log-likelihood. For Gaussian mixture model-based clustering, the complete-data log-likelihood is
\begin{equation}\label{eqn:cdll}
\log\mathcal{L}_{\text{c}}(\varthet\mid\vecx_1,\ldots,\vecx_n,\vecz_1,\ldots,\vecz_n)=\sum_{i=1}^n\sum_{g=1}^Gz_{ig}\log[\pi_g\phi(\vecx_i\mid\vecmu_g,\matsig_g)].
\end{equation}
The M-step involves maximizing the expected value of Equation~\ref{eqn:cdll} with respect to the model parameters. The E- and M-steps are iterated until convergence. Details on the EM algorithm parameter estimates for the GPCMs are given by \cite{celeux95} and \cite{fraley99}. 

One feature of these EM algorithms is the importance of starting values: {\tt mclust} utilizes a Gaussian model-based agglomerative hierarchical clustering procedure to obtain starting values \citep[cf.][]{murtagh84,banfield93}. The {\tt mixture} package allows the user freedom in selecting starting values, with $k$-means clustering \citep{hartigan79} results being the default.
The clusterings for a given model arise as the maximum \textit{a~posteriori} (MAP) expected values (i.e., probabilities) of the $Z_{ig}$. To compute these MAP values, we compute the expected values  
\begin{equation}
\mathbb{E}[Z_{ig}\mid\hat{\varthet}] = \frac{\hat{\pi}_g\phi(\vecx_i\mid\hat{\vecmu}_g,\hat{\matsig}_g)}{ \sum_{h=1}^G\hat{\pi}_h\phi(\vecx_i\mid\hat{\vecmu}_h,\hat{\matsig}_h)}\equalscolon\hat{z}_{ig},
\end{equation}
with the parameter estimates $\hat{\varthet}$ taking the converged values. Then, $\text{MAP}\{\hat{z}_{jg}\}=1$ if max$_g\{ \hat{z}_{jg} \}$ occurs at component $g$ and $\text{MAP}\{\hat{z}_{jg}\}=0$ otherwise. 
Note that the expected values $\hat{z}_{ig}$ are computed in each E-step, which is why we emphasize that in computation of the MAP classifications $\hat{z}_{ig}$ depends on the parameter values at convergence.

The Bayesian information criterion \citep[BIC;][]{schwartz78} is used to select the number of components and the covariance structure (i.e., the model). Although it is by far the most popular model selection criterion for mixture model-based clustering, the regularity conditions for the asymptotic approximation used in the development of the BIC are not generally satisfied by mixture models \cite[cf.][]{keribin98,keribin00}. There is, however, plenty of practical evidence to support its use in mixture model selection \citep[e.g.,][]{dasgupta98,fraley02a} and we use it for the analyses herein. The BIC is given by $-2l(\vecx, \hat{\varthet})+m\log n$,
where $m$ is the number of free parameters, 
$l(\vecx,\hat{\varthet})$ is the maximized log-likelihood, and $\hat{\varthet}$ is the maximum likelihood estimate of $\varthet$. \cite{dasgupta98} proposed using the BIC for mixture model selection, where the model with the lowest BIC is selected.

\section{Methodology}\label{sec:method}
\subsection{Constrained Covariance Updates} \label{constrained updates}

Our constrained EM algorithm is an alternating conditional maximization algorithm \citep{meng97}, where the matrix of eigenvalues $\matb_g$ for each group is maximized conditional on $\matd_g$ and then \textit{vice versa}. Note, that $\matb_g$ or $\matd_g$ can be equal or varying across groups, depending on the model (cf. Table~\ref{tab:mclust3}). These conditional updates can be repeated $m$ times or until a convergence criteria is achieved.  \cite{ingrassia2007,ingrassia2011} give an algorithm for our VV model when the smallest eigenvalue is constrained and show that these constraints maintain the monotonicity property. Herein, we introduce and illustrate parameter estimation with constraints on both the smallest and largest eigenvalues. 

Let $[a,b]$ be the range of allowable eigenvalues and let $\mats_g$ be the sample covariance matrix for group $g$, i.e., $\sampcov_g=(1/n_g)\sum_{i=1}^{n}\hat{z}_{ig}(\vecx_i-\vecmu_g)(\vecx_i-\vecmu_g)'$. 
%
Consider unconstrained $\matb_g$. From \cite{celeux95}, we have 
\begin{equation*}
\vecv^{(t+1)}_g = \mbox{diag}\left\{\matd_g^{(t)} \mats_g \matd_g^{(t)}\right\},
\end{equation*}
where $\vecv^{(t+1)}_g = ( v_{g1}^{(t)}, \ldots, v_{gp}^{(t)} )$ and the superscripts in parentheses denote iteration number. If we let $b_{g1},\ldots, b_{gp}$ be the diagonal elements of $\matb^{(t+1)}_g $, then the constrained EM uses the updates 
\begin{equation*}
b_{gk}^{(t+1)} = \mbox{min}\left\{b, \mbox{max}\left( v_{gk}^{(t+1)}, a\right)\right\}.
\end{equation*}


Now, suppose we set $\matb_g = \matb$. Then,
\begin{equation*}
\vecv^{(t+1)} = \mbox{diag}\left\{\sum_g^G \pi_g^{(t+1)} \matd_g^{(t)} \mats_g \matd_g^{(t)}\right\},
\end{equation*}
where $\vecv^{(t+1)} = ( v_{1}^{(t+1)}, \ldots, v_{p}^{(t+1)} )$.
If we let $b_{1},\ldots, b_{p}$ be the diagonal elements of $\matb^{(t+1)} $, then the constrained EM algorithm sets 
\begin{equation*}
b_{gk}^{(t+1)} = \mbox{min}\left\{b,\mbox{max}\left(v_k^{(t+1)},a\right)\right\}.
\end{equation*}

Consider unconstrained $\matd_g$ and 
let  $\mats_g = \matp_g\matq_g \matp'_g $ be the eigen-decomposition of $\mats_g$. Then we set  
$\matd_g^{(t+1)} = \matp_g.$

Finally, consider $\matd_g = \matd$. This update can be carried out using Flury's method \citep[see][for details]{flury1984,celeux95}.

\subsection{Dynamic  Initialization}

We run the EM algorithm for each member of the rGPCM family as described in Section~\ref{constrained updates}, but for the first $k$ iterations we use a sequence of $k$ constraints $S = \left\{ (a_1,b_1), \ldots, (a_k,b_k)  \right\}$.  We could use such a set $S$; however, we instead simplify and use a sequence $v = \left\{ 0, \ldots, 1\right\} $, where $v$ is some sequence from $0$ to $1$. We also use the mapping
\begin{equation}
(a_i,b_i) = \left( a(v_i), b(v_i) \right) = \beta \left( 1-v_i   ,   1-\log(1-v_i)      \right),
\end{equation}
where $\beta >0$. 
These equations are set up so that $v_1=0$ and $v_k=1$ implies that $(a_1,b_1) = (
\beta,\beta)$ and $(a_k,b_k) = (0, \infty)$. In this paper, we have set $\beta=1$  because we scale the data in our clustering applications (Section~\ref{sec:anal}).

\section{Data Experiments} \label{sec:anal}

\subsection{Performance Assessment}

Although our examples are all genuine clustering problems, i.e., no knowledge of labels are used, the labels are known in each case; therefore, we can asses the performance of our algorithms for the rGPCM family. We use classification tables and adjusted Rand indices \citep[ARI;][]{hubert85} to summarize classification accuracy. The ARI corrects the Rand index \citep{rand71} for chance agreement. An ARI value of 1 indicates perfect class agreement and a value of 0 would be expected under random classification.

\subsection{Simulation Study 1}
\begin{figure}[!htp]
\begin{center}
\includegraphics[width=0.65\textwidth,angle=270]{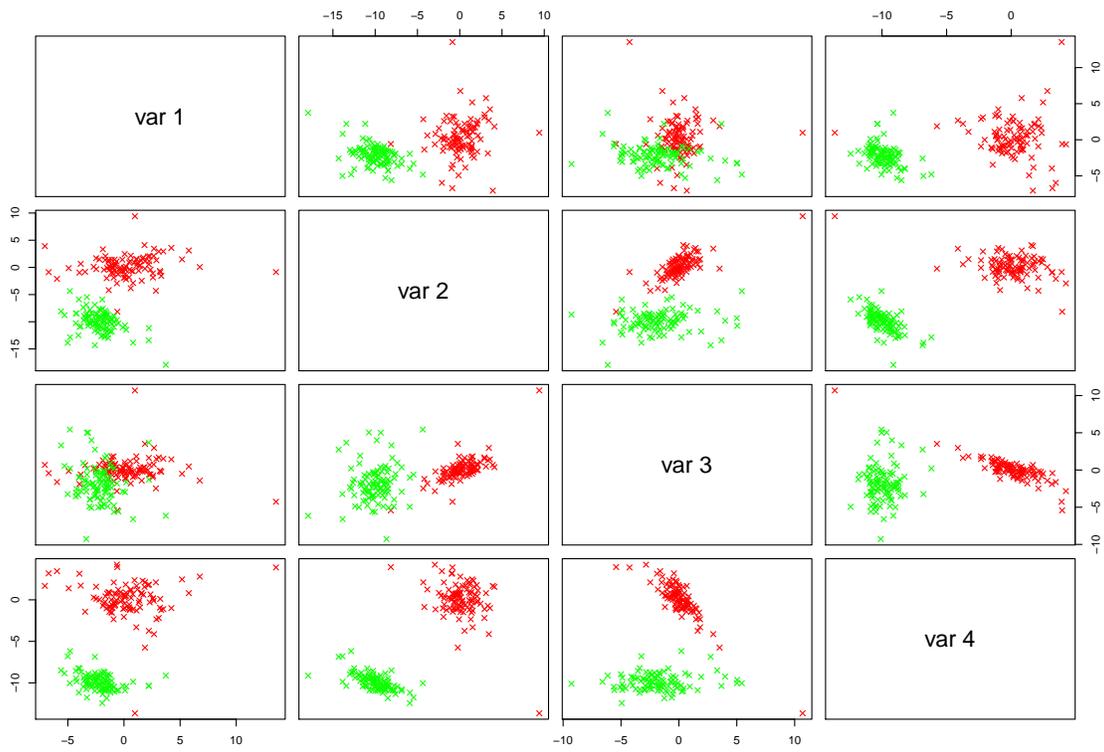}
\caption{\label{sim1}Simulated data from a four-dimensional two-component EV model with heavy tails.}
\end{center}
\end{figure}

The first data set consists of $n=200$ observations generated from a four-dimensional two-component ($n_1=100$ and $n_2=100$) mixture of multivariate $t$-distributions with scale matrix $\matsig_g=\lambda\matd_g\mata\matd_g'=\matd_g\matb\matd_g'$ and 5 degrees of freedom. As we would expect, the resulting clusters are clearly heavy tailed with several outlying points (Figure~\ref{sim1}). Using the output from $k$-means clustering as the initialization for $\mathbf{z}_1,\ldots,\mathbf{z}_{n}$, we run our algorithm for $G=1,\dots,6$. The eigenvalues are constrained to be within the smallest and largest eigenvalues of the sample covariance matrix of the data. 

\begin{figure}[!htp]
\begin{center}
\includegraphics[width=0.65\textwidth,angle=270]{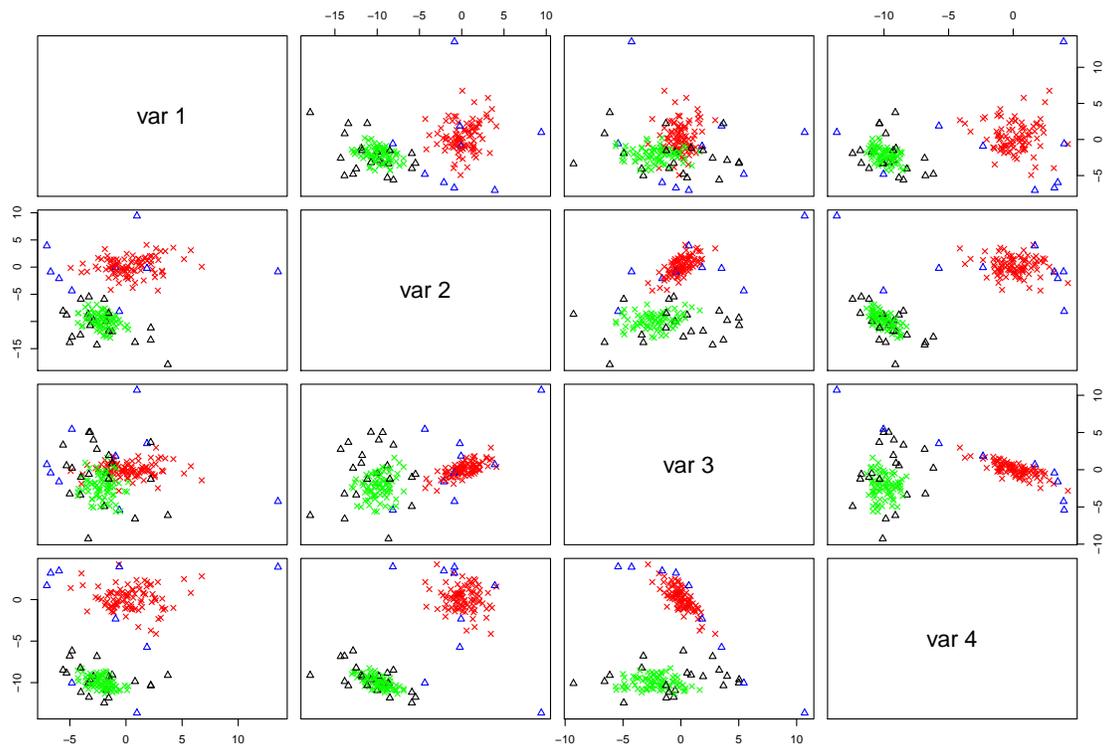}
\caption{\label{mclusts1} Predicted classifications using the best GPCM model for simulation study~1.}
\end{center}
\end{figure}

The {\tt mixture} package is used to fit the GPCM models to facilitate comparison with the rGPCM family. The chosen rGPCM model is a two-component EV model that gives perfect classification ($\text{ARI}=1$), whereas the selected GPCM model is a four-component VEV ($\matsig_g=\lambda_g\matd_g\mata\matd_g'$) model with $\text{ARI}=0.75$ (Table~\ref{tabsim1}). In the absence of a constraint on eigenvalues, the chosen GPCM model has additional components with relatively high variance to accommodate the heavier tails (Figure~\ref{mclusts1}). From Table~\ref{tabsim1}, it is clear that with appropriate merging of components, the classification performance of the best GPCM model is very close to that of the best rGPCM model. BIC values for all rGPCM models are given in Appendix~\ref{app:secbic}.
\begin{table}[!h]
\caption{\label{tab:sim1p} Classifications for the best rGPCM and GPCM models, respectively, for simulation study~1.}
\centering
\begin{tabular*}{0.95\textwidth}{@{\extracolsep{\fill}}l|lllllll}
\hline
 {True $\backslash$ Estimated}  & \multicolumn{2}{c}{rGPCM} && \multicolumn{4}{c}{GPCM} \\
\cline{2-3} \cline{5-8}
&1&2&&1&2&3&4\\
\hline
1&100&&&92&8& & \\
   2& & 100 && &1&  21& 78\\
\hline\end{tabular*}\label{tabsim1}
\end{table}

\subsection{Simulation Study 2}
\begin{figure}[!htp]
\centering
\subfigure[Without noise.]{
\includegraphics[width=0.45\textwidth,angle=270]{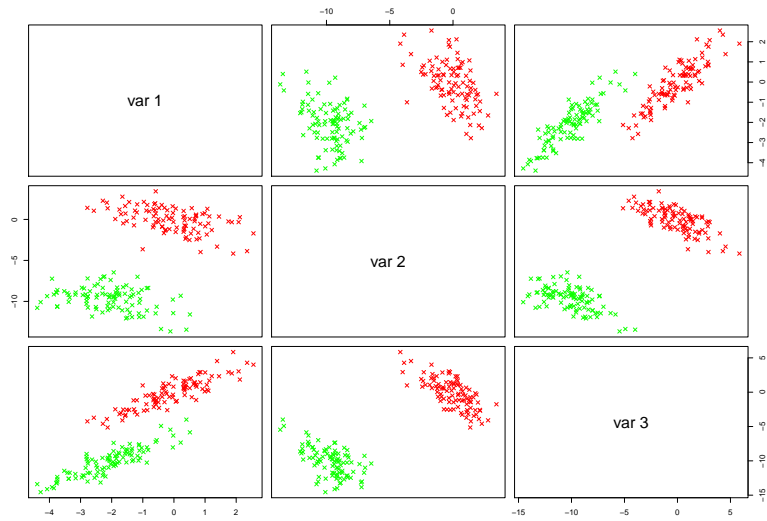}\label{no-noise}
}
\quad
\subfigure[ With 5\% uniform noise.]{
\includegraphics[width=0.455\textwidth,angle=270]{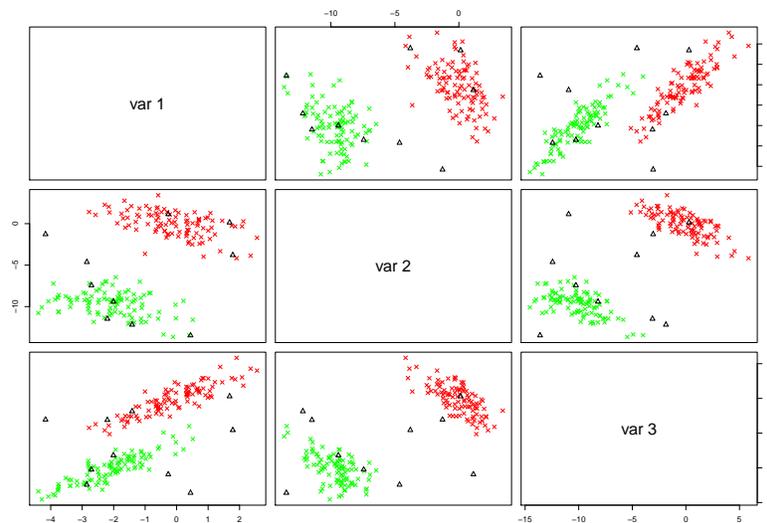}\label{noise}
}
\caption{
Simulated data from simulation study 2.
}
\end{figure}

The second data set consists of $n=200$ observations generated from a three-dimensional two-component Gaussian mixture model with $\matsig_g=\lambda\matd\mata\matd'=\matd\matb\matd'$, $n_1=100$, and $n_2=100$. As shown in Figure~\ref{no-noise}, the components are very well separated in these simulated data. 
%
Again, we run our algorithms for the rGPCM models for $G=1,\ldots,6$ with the eigenvalues constrained to lie within the smallest and largest eigenvalues of the sample covariance matrix. 
For the same data, we also run the GPCM models for $G=1,\ldots,6$. Both algorithms selected a $G=2$ component EE model and give perfect classifications. BIC values for the rGPCM models are given in Appendix~\ref{app:secbic}.

We then added 5\% uniform noise to the data set ($n_3=10$ noise observations, see Figure~\ref{noise}) and repeated the above analyses for the rGPCM and the GPCM families, respectively. The selected rGPCM model is a two-component EE model that absorbed the noisey observations into the two Gaussian components (Table~\ref{tabsim2}). On the other hand, the chosen GPCM model is a eight-component EEE model ($\matsig_g=\lambda_g\matd_g\mata\matd_g'$), where four of the components contain just one or two of the noisy points and the other two components contain the true Gaussian components along with one and three noisy points, respectively (Table~\ref{tabsim2}). Again, BIC values for the rGPCM models are given in Appendix~\ref{app:secbic}. 
\begin{table}[!h]
\caption{\label{tab:sim1p} Classifications for the best rGPCM and GPCM models, respectively, for simulation study~2.}
\centering
\begin{tabular*}{0.95\textwidth}{@{\extracolsep{\fill}}l|lllllllll}
\hline
 {True $\backslash$ Estimated}  & \multicolumn{2}{c}{rGPCM} && \multicolumn{6}{c}{GPCM} \\
\cline{2-3} \cline{5-10}
&1&2&&1&2&3&4&5&6\\
\hline
1&100& & &100&&&&& \\
2& &100 && &&&&&100\\
3&4&6&& 1&2&2&1&1&3\\
\hline\end{tabular*}\label{tabsim2}
\end{table}

We have illustrated the effect of only a very small proportion of outliers on the EM algorithms used for parameter estimation for the GPCM family versus the application of our algorithms for the rGPCM family. Please note that we are not proffering the rGPCM solution as being ideal in this context; rather, we are suggesting that our parameter estimation approach led to rGPCM results that are preferable to those obtained using the more traditional parameter estimation approach for the GPCM family. Effective methods for clustering noisy data include trimmed clustering \citep[e.g.,][]{garcia08,garcia10} and mixtures of contaminated distributions \citep{punzo13}. 

\subsection{Two Well-Known Data Sets}

\cite{forina86} recorded 28 chemical and physical properties of three types of wine (Barolo, Grignolino, Barbera) from the Piedmont region of Italy. A subset of 13 of these variables is available in the {\tt gclus} package \citep{hurley04} for {\sf R}. 
The {\em leptograpsus} crabs data set can be found in the {\tt MASS}  package \citep{venables99} in {\sf R}. These data contain five physical measurements on two different colours of crab, further separated into gender. MCLUST is known to do poorly on these data; \cite{raftery06} used these data to illustrate the superiority of their variable selection technique over MCLUST. 

\subsection{Illustrating Convergence From Random Starting Values}

For each data set, we generate 50 random starting points. We run the four types of EM algorithm until convergence for  $G=2,\ldots,6$ components. Specifically, we run the EM algorithm in four circumstances: no constraints, lower constraints, upper constraints, and both upper and lower (range) constraints on the eigenvalues. For each dynamic initialization, we use an equidistant sequence of length 25 from $0$ to $1$. 
For each run, we noted which algorithms achieved the highest converged log-likelihood value for a particular starting value. This is because all four algorithms could, and sometimes did, converge to the same solution. 

The results are given Tables~\ref{A1} to~\ref{A4} in Appendix~\ref{app:tabs}. By inspection of these tables, the value of imposing eigenvalue constraints is clear. Specifically, the model most often converges to the `best' value of the log-likelihood is very rarely from an unconstrained EM algorithm. Furthermore, the unconstrained EM algorithm yields far more degenerate solutions that its constrained counterparts.


\subsection{Constrained Eigenvalues: A Comparison With The GPCM Family}

For each data set, we compare results for the rGPCM models using the constrained eigenvalue approach to parameter estimation  to results for the GPCM models with the traditional EM algorithm approach to parameter estimation. 
When estimating parameters for the rGPCM family, we constrain eigenvalues to be within the smallest and largest eigenvalues of the sample covariance matrix of each data set, i.e., $[0.1033,4.7057]$ for the wine data and $[0.0017,4.7888]$ for the crabs data. BIC values for the rGPCM models for all data sets are given in Appendix~\ref{app:secbic}.

For the wine data, the best rGPCM model is a $G=3$ component VE model with an ARI of 0.96 (Table~\ref{w2}). The best GPCM model is a $G=3$ component model with an ARI of 0.90.
\begin{table}[!h]
\caption{\label{w2}Classification table associated with the best GPCM model for the wine data.}
{\begin{tabular*}{1.0\textwidth}{@{\extracolsep{\fill}}l|rrr}
\hline
 &  1 &  2 &  3 \\
\hline  
Barolo& 58&  1 &\\
Grignolino&  1& 70& \\
Barbera&  &  & 48 \\
\hline
\end{tabular*}}
\end{table}

For the crabs data, the best rGPCM model is a $G=4$ component EV model with an ARI of 0.80 (Table~\ref{c2}). For the crabs data, the best GPCM model is a $G=9$ component model with an ARI of 0.50.
\begin{table}[!h]
\caption{\label{c2}Classification table for the best rGPCM model for the crabs data.}
{\begin{tabular*}{1.0\textwidth}{@{\extracolsep{\fill}}l|rrrr}
\hline
 & 1&  2&  3&  4 \\
\hline
   Blue \& Male & 38&  12& & \\
   Blue \& Female&  & 50& & \\
 Orange \& Male& & &  50&\\
 Orange \& Female& & & 5&45 \\
 \hline
\end{tabular*}}
\end{table}

\section{Discussion} \label{sec:conc}

In this paper, we introduced a constrained eigenvalue parameter estimation procedure for the eight of the parsimonious Gaussian clustering models of \cite{celeux95}. For convenience, we have referred to this subset of models as the rGPCM family. Please note that when we discuss the performance of the rGPCM family herein, we are referring to the performance of those models with our constrained eigenvalue parameter estimation procedure. We are not suggesting that the rGPCM models are in any sense better than the other GPCM models when the same parameter estimation methods are used.

We illustrated our approach through extensive simulation studies and two real data applications. In one application, we studied dynamic initialization, where we begin with random starting values and impose stringent constraints on the eigenvalues which are slowly lifted during the first $25$ iterations of the EM algorithm. This approach is shown to maintain the monotonicity of the EM algorithm while reducing the risk of converging to a degeneracy. 
In another application, we constrained the range of eigenvalues and fit the rGPCM models to two well known data sets. In most cases, we find solutions that are an improvement over the famous GPCM models; however, we require constraints to be chosen \emph{a priori}. Constraining the eigenvalues in this way can be viewed as a form of regularization or as placing a uniform prior on the eigenvalues. Future work will involve studying different approaches to estimating the range of allowable eigenvalues.

The fact that the rGPCM models outperformed the GPCM models on both simulation studies and real data sets shows that the eigenvalue constraints we use can lead to improved classification performance. Furthermore, if one follows the approach of only running the rGPCM models for which our eigenvalue constraints can be used, this is tantamount to merging the volume ($\lambda$ or $\lambda_g$) and shape ($\mata$ or $\mata_g$) parameters from the famous eigen-decomposition used in the GPCM family. Therefore, the importance of having separate volume and shape parameters deserves further consideration.  Furthermore, even if it can be useful in some scenarios, the value of including the component volume as a separate parameter has to be judged in context with the fact that including it prevents application of the constrained eigenvalue approach to parameter estimation. 


\appendix

\section{BIC Tables}\label{app:secbic}
\begin{table}[!htp]
\caption{\label{tabbicsim1} BIC values for our constrained eigenvalue models for simulation study~1, where NA entries indicate that a model failed to converge.}
{\scriptsize\begin{tabular*}{1.0\textwidth}{@{\extracolsep{\fill}}c|cccccccc}
\hline
 $G$&   EI &   VI  &  EE  &  EV &   VV &   VE  &  GI  &  1I\\
 \hline
1& $838.4$ & $838.4$ &$600.1$ &$514.5$& $514.5$ &$600.1$& $822.5$& $822.5$\\
2& $206.7$ &$173.1$ &$176.7$ & $\mathbf{78.5}$  &$87.6$&$143.4$ &$474.6$ &$499.6$\\
3 &$144.0$ & $90.5$ &$149.3$ & $82.8$ &$107.7$ & $86.6$ &$227.1$& $431.3$\\
4 &$129.4$ &  NA &$110.1$ &$124.8$ &  NA  &  NA  &  NA&$403.5$\\
5 &$113.3$ &  NA & $95.7$ &$143.1$  &  NA  &  NA   & NA &$347.4$\\
6 &$122.4$  &  NA &$104.1$ &$183.2$ &  NA  &  NA  &  NA &$293.9$\\
\hline
\end{tabular*}}
\end{table}

\begin{table}[!htp]
\caption{\label{tabbicsim2} BIC values for our constrained eigenvalue models for simulation study~2.}
{\scriptsize\begin{tabular*}{1.0\textwidth}{@{\extracolsep{\fill}}c|cccccccc}
\hline
 $G$&   EI &   VI  &  EE  &  EV &   VV &   VE  &  GI  &  1I\\
 \hline
1 &  $659.1$& $659.1$ &  $219.4$  & $205.1$ &  $205.1$ & $219.4$ & $648.4$& $648.4$\\
2 &   $14.3$ &  $28.7$ &  $\mathbf{-172.3}$ & $-156.9$ & $-141.9$ & $-157.4$ & $172.4$ & $167.2$\\
3 &  $-27.9$ & $-11.5$ &  $-162.4$ & $-143.0$ & $-123.8$ & $-153.1$ &$74.7$ & $98.4$\\
4 &  $-84.1$ & $-41.6$ & $-154.7$ & $-119.5$ & $-97.5$ & $-138.9$ & $3.8$ & $-4.8$\\
5 &  $-83.7$ & $-27.9$ & $-134.3$  & $-82.6$ & $-41.6$& $-102.8$ & $-26.6$& $-28.4$\\
6& $-102.2$ & $-38.5$ & $-105.7$ &  $-67.2$ & $6.1$ & $-73.4$ & $-67.6$ & $-73.0$\\
\hline
\end{tabular*}}
\end{table}

\begin{table}[!h]
\caption{\label{w1}BIC values for our constrained eigenvalue models applied to the wine data.}
{\scriptsize\begin{tabular*}{1.0\textwidth}{@{\extracolsep{\fill}}c|cccccccc}
\hline

$G$&     EI    & VI  &   EE  &   EV   &  VV   &  VE &    GI  &   1I\\
\hline  
1& 2435.7& 2435.7 &2071.5 &1475.4 &1475.4 &2071.5 &2373.5 &2373.5\\
2 &1808.8 &1717.8 &1414.5 &1424.6 &1420.3 &1245.0 &1830.9 &1903.5\\
3 &1399.9 &1291.3 &1311.4 &1580.8 &1642.7& \textbf{1164.6} &1442.9 &1526.8\\
4 &1353.6 &1280.5 &1291.4& 1925.7& 1992.2 &1201.9 &1439.7 &1495.3\\
\hline
5 &1332.4 &1371.9 &1256.3& 2207.7 &2401.6 &1218.6 &1400.6 &1470.4\\
6 &1297.1 &1354.9 &1328.7 &2509.2 &2854.1 &1348.1 &1368.4 &1463.9\\
7 &1283.2 &1394.1& 1284.0 &2965.4   &  NA& 1334.2 &1395.4 & 1422.0\\
8 & 1359.7 &1488.5 &1373.5& 3257.6  &   NA &1410.6& 1443.3& 1544.5\\
9 &1255.9& 1502.1& 1368.9 &3631.1  &   NA &1443.3 &1416.9 &1446.6\\
\hline
\end{tabular*}}
\end{table}

\begin{table}[]
\caption{\label{c1}BIC values for our constrained eigenvalue models applied to the crabs data.}
{\scriptsize\begin{tabular*}{1.0\textwidth}{@{\extracolsep{\fill}}c|cccccccc}
\hline
 $G$ &   EI  &   VI   &   EE   &   EV   &   VV  &    VE  &   GI    & 1I\\
\hline  
1 &  1048.0 & 1048.0 &  -506.3 & -1749.6 & -1749.6  & -506.3 & 1026.8&  1026.8\\
2 &   304.8 &  328.5 & -1784.3 & -1756.7 & -1756.3 & -1786.6 &  288.3 &  285.1\\
3  & -101.0 &  -89.1 & -1727.3 & -1909.9 & -1894.8&  -1742.2 & -129.1&  -101.6\\
4  & -423.6 & -392.9 & -1811.3 & \textbf{-1963.9} & -1595.3 & -1705.0 & -456.4 & -423.3\\
5  & -642.7 & -492.8 & -1834.5 & -1724.8 & -1595.8 & -1812.5 & -610.2 & -579.6\\
6  & -743.2 & -722.1 & -1949.0 & -1698.1 & -1532.7 & -1743.5 & -643.4 & -620.3\\
\hline 
7 &  -860.5&  -773.8 & -1923.7 & -1610.0 & -1417.2 & -1715.9&  -734.2 & -716.5\\
8 &  -883.4 & -684.2 & -1892.0 & -1558.7 & -1368.9 & -1716.3&  -664.3 & -739.0\\
9  & -960.8 & -792.7 & -1770.9 & -1445.7 & -1381.5 & -1653.3 & -798.7 & -738.0\\
10&  -898.2&  -817.8 & -1913.9 & -1412.5 & -1267.3 & -1642.9 & -829.7 & -787.3\\
11 & -938.9 & -729.1&  -1897.8 & -1427.9 & -1159.0 & -1641.6 & -841.4 & -743.2\\
12 & -924.1 & -725.9 & -1868.7 & -1292.8 & -1065.9 & -1552.8 & -863.5 & -882.3\\
\hline
\end{tabular*}}
\end{table}
\newpage

\section{Convergence Tables}\label{app:tabs}

\begin{table}[]
\caption{\label{A1}The proportion of times each algorithm converged to the highest log-likelihood, given a particular starting value, on the wine data set.}
{\scriptsize\begin{tabular*}{1.0\textwidth}{@{\extracolsep{\fill}}cc|cccc}
\hline
 & &  \multicolumn{4}{c}{Constraint Type}\\
 \cline{4-5}
Model & $G$ & None & Lower & Upper & Range\\
\hline  
EI& 2& 1.00& 1.00& 1.00& 1.00 \\
EI& 3& 1.00& 1.00& 1.00& 1.00 \\
EI& 4& 1.00& 1.00& 1.00& 1.00 \\
EI& 5& 1.00& 1.00& 1.00& 1.00 \\
EI& 6& 1.00& 1.00& 1.00& 1.00 \\
\hline
VI& 2& 1.00& 1.00& 1.00& 1.00 \\
VI& 3& 1.00& 1.00& 1.00& 1.00 \\
VI& 4& 1.00& 0.96& 1.00& 0.96 \\
VI& 5& 0.96& 1.00& 0.96& 1.00 \\
VI& 6& 0.88& 1.00& 0.88& 1.00 \\
\hline
EE& 2& 0.62& 0.52& 0.42& 0.36 \\
EE& 3& 0.68& 0.58& 0.74& 0.70 \\
EE& 4& 0.86& 0.66& 0.84& 0.76 \\
EE& 5& 0.64& 0.68& 0.56& 0.46 \\
EE& 6& 0.68& 0.52& 0.62& 0.56 \\
\hline
EV& 2& 0.88& 0.70& 0.88& 0.68 \\
EV& 3& 0.44& 0.60& 0.44& 0.82 \\
EV& 4& 0.48& 0.82& 0.42& 0.76 \\
EV& 5& 0.24& 0.52& 0.24& 0.36 \\
EV& 6& 0.22& 0.40& 0.14& 0.46 \\
\hline
VV& 2& 0.82& 0.90& 0.74& 0.76 \\
VV& 3& 0.38& 0.68& 0.38& 0.62 \\
VV& 4& 0.18& 0.58& 0.08& 0.44 \\
VV& 5& 0.16& 0.54& 0.16& 0.30 \\
VV& 6& 0.10& 0.62& 0.02& 0.38 \\
\hline
VE& 2& 0.98& 0.90& 0.90& 0.84 \\
VE& 3& 0.64& 0.40& 0.60& 0.54 \\
VE& 4& 0.36& 0.62& 0.26& 0.82 \\
VE& 5& 0.28& 0.32& 0.46& 0.26 \\
VE& 6& 0.20& 0.34& 0.30& 0.32 \\
\hline
\end{tabular*}}
\end{table}


\begin{table}[]
\caption{\label{A4}The proportion of times each algorithm converged to the highest log-likelihood and degeneracy occurred, given a particular starting value, on the crabs data set.}
{\scriptsize\begin{tabular*}{1.0\textwidth}{@{\extracolsep{\fill}}cc|cccc|cccc}
\hline
 & &  \multicolumn{4}{c|}{Log-likelihood} & \multicolumn{4}{c}{\% Degeneracy}\\
\cline{4-5} \cline{8-9}
Model & $G$ & None & Lower & Upper & Range& None & Lower & Upper & Range\\
\hline  
EI& 2& 1.00& 1.00& 1.00& 1.00& 0.00& 0.00& 0.00& 0.00 \\
EI& 3& 1.00& 0.96& 1.00& 0.96& 0.00& 0.00& 0.00& 0.00 \\
EI& 4& 1.00& 0.96& 1.00& 0.96& 0.00& 0.00& 0.00& 0.00 \\
EI& 5& 0.96& 0.84& 0.96& 0.84& 0.00& 0.00& 0.00& 0.00 \\
EI& 6& 0.96& 0.84& 0.96& 0.84& 0.00& 0.00& 0.00& 0.00 \\
\hline
VI& 2& 1.00& 1.00& 1.00& 1.00& 0.00& 0.00& 0.00& 0.00 \\
VI& 3& 0.96& 1.00& 0.96& 0.96& 0.00& 0.00& 0.00& 0.00 \\
VI& 4& 0.76& 0.84& 0.76& 0.88& 0.00& 0.00& 0.00& 0.00 \\
VI& 5& 0.60& 0.68& 0.64& 0.64& 0.04& 0.04& 0.04& 0.04 \\
VI& 6& 0.56& 0.36& 0.68& 0.44& 0.00& 0.08& 0.04& 0.04 \\
\hline
EE& 2& 0.00& 0.00& 1.00& 1.00& 0.00& 0.00& 0.00& 0.00 \\
EE& 3& 0.04& 0.12& 0.96& 1.00& 0.00& 0.00& 0.00& 0.00 \\
EE& 4& 0.04& 0.48& 0.88& 0.92& 0.00& 0.00& 0.00& 0.00 \\
EE& 5& 0.08& 0.52& 0.56& 0.56& 0.00& 0.00& 0.00& 0.00 \\
EE& 6& 0.00& 0.56& 0.52& 0.16& 0.00& 0.00& 0.00& 0.00 \\
\hline
EV& 2& 0.12& 1.00& 1.00& 1.00& 0.00& 0.00& 0.00& 0.00 \\
EV& 3& 1.00& 0.84& 0.96& 1.00& 0.00& 0.00& 0.00& 0.00 \\
EV& 4& 0.48& 0.20& 0.32& 0.44& 0.00& 0.00& 0.00& 0.00 \\
EV& 5& 0.48& 0.04& 0.24& 0.24& 0.00& 0.00& 0.00& 0.00 \\
EV& 6& 0.36& 0.20& 0.28& 0.32& 0.00& 0.00& 0.00& 0.00 \\
\hline
VV& 2& 0.60& 1.00& 1.00& 1.00& 0.00& 0.00& 0.00& 0.00 \\
VV& 3& 0.28& 0.48& 0.60& 0.68& 0.00& 0.20& 0.04& 0.00 \\
VV& 4& 0.12& 0.52& 0.44& 0.52& 0.12& 0.20& 0.04& 0.08 \\
VV& 5& 0.12& 0.24& 0.28& 0.52& 0.32& 0.48& 0.16& 0.04 \\
VV& 6& 0.12& 0.20& 0.28& 0.28& 0.52& 0.68& 0.36& 0.24 \\
\hline
VE& 2& 1.00& 1.00& 1.00& 1.00& 0.00& 0.00& 0.00& 0.00 \\
VE& 3& 0.24& 0.56& 0.88& 0.92& 0.00& 0.00& 0.00& 0.00 \\
VE& 4& 0.12& 0.24& 0.28& 0.64& 0.04& 0.20& 0.16& 0.12 \\
VE& 5& 0.44& 0.28& 0.40& 0.24& 0.00& 0.36& 0.16& 0.08 \\
VE& 6& 0.32& 0.12& 0.44& 0.12& 0.08& 0.60& 0.20& 0.24 \\
\hline
\end{tabular*}}
\end{table}

\end{document}